\documentclass[amsmath,amssymb,nofootinbib]{revtex4}
\usepackage{color}
\usepackage{graphicx}
\usepackage{subfigure}

\begin{document}

\title{Gluino bounds: Simplified Models vs a Particular SO(10) Model\\ (A Snowmass white paper)}

\author{Archana Anandakrishnan}
\author{B. Charles Bryant}
\author{Stuart Raby}
\affiliation{Department of Physics, The Ohio State University,
191 W.~Woodruff Ave, Columbus, OH 43210, USA}

\author{Ak\i{}n Wingerter}
\affiliation{Laboratoire de Physique Subatomique et de
Cosmologie,UJF Grenoble 1, CNRS/IN2P3, INPG,53 Avenue des Martyrs, F-38026 Grenoble,France}

\begin{abstract}
We consider the results from the first run of LHC studied in the
context of simplified models and re-interpret them for a particular SO(10) model with a non-simplified topology.
Hadronic searches have been designed to obtain the best sensitivity for the simplified models.  They require
multiple b-jets in the final state.  But we show that the bounds obtained from these searches are weaker in the case of the particular model studied here, since there are fewer b-jets.
\end{abstract}

\maketitle

The results from the first run of the LHC have placed significant bounds on
supersymmetric models. In order to present the results in a maximally
model-independent manner, LHC uses the framework of simplified models \cite{Alwall:2008ag}.
Simplified models are described by a small number of parameters and can be used
to obtain constraints on a wide range of models. Even in the case of general models,
simplified models can be used to place upper limits on the production cross sections.
We were motivated to contrast the bounds obtained for simplified models against a particular
model with a very non-simplified topology. In Ref. \cite{Anandakrishnan:2013nca}, we studied
benchmark models from an SO(10) SUSY GUT with Yukawa unification. \\

We considered 15 benchmark models with varying gluino masses between 300 GeV to 2 TeV based on fits to
low energy observables. In minimal models of Yukawa unification with ``Just-so splitting'' for the Higgs masses
at the GUT scale, the gluino prefers to be light. For universal scalar masses, $m_{16} = 20$ TeV at the GUT scale, the
best fit gluino mass is around 600 GeV. A general feature of the model is that the squarks and sleptons of
the first two families have a large mass of order $m_{16}$ which are constrained to be greater than 10 TeV, based on
fits to flavor physics observables.
The third family of squarks and sleptons, however,
are significantly lighter (few TeV) and finally the gauginos are even lighter.
Therefore, the simplified models which might come closest to constraining our
results are those in which gluinos are pair produced at the LHC and then decay
with 100\% branching ratio  $\tilde g \rightarrow t \ \bar t \ \tilde \chi^0_1$ (T1tttt)
or $\tilde g \rightarrow b \ \bar b \ \tilde \chi^0_1$ (T1bbbb).  The lower bound on the
gluino mass from CMS \cite{Chatrchyan:2013wxa,Chatrchyan:2012paa,Chatrchyan:2013lya} and ATLAS
\cite{ATLAS-CONF-2013-007,ATLAS-CONF-2013-054,ATLAS-CONF-2013-047} searches are $M_{\tilde g} \gtrsim 1 - 1.2$ TeV,  where the actual bound
depends on the search strategy.\\

To illustrate the main results of our analysis, we present a sample benchmark model,
(YUc in Ref. \cite{Anandakrishnan:2013nca}) with a gluino mass of 1 TeV in TABLE \ref{benchmark}.
The spectrum illustrates an inverted mass hierarchy for the scalars, and the decay topology is not simplified.
Since the neutral Higgsinos are lighter than the gluino, loop decays of the gluino to a gluon and a neutralino are enhanced.\\

\begin{table}[h!]
\begin{tabular}{|l|l r |l r |l r |}
\hline
GUT scale parameters & $m_{16}$ & 20 & $M_{1/2}$ & 0.25  & $A_0$  & -41  \\
\hline
EW parameters & $\mu$ & 0.8  & tan$\beta$ & 50 & & \\
\hline
Spectrum & $m_{\tilde t_1}$ & 3.695  & $m_{\tilde b_1}$  & 4.579  & $m_{\tilde \tau_1}$  & 7.834  \\
& $m_{\tilde\chi^0_1}$   & 0.172  & $m_{\tilde\chi^+_1}$  &  0.342  & $M_{\tilde g}$   & 1.061   \\
& $m_{{\tilde u},{\tilde d},{\tilde e}}$ & 20 & $m_{{\tilde c},{\tilde s},{\tilde \mu}}$ & 20 &  $M_A$ & 2.2 \\
\hline
Gluino Branching Fractions & $ g \widetilde{\chi}^0_4$ & 38\% & $ g \widetilde{\chi}^0_3$& 35\%& $
tb\widetilde{\chi}^\pm_1$ & 14\% \\
& $ g \widetilde{\chi}^0_2$ & 8\% &$t\bar{t}\widetilde{\chi}^0_1$& 1.2\% & $ b\bar{b}\widetilde{\chi}^0_1$ & 0.006\% \\
\hline
\end{tabular}
 \caption{\label{benchmark} Benchmark model YUc from Ref. \cite{Anandakrishnan:2013nca}. All masses are in TeV.}
\end{table}

From a wide range of available experimental analyses, we picked those
that are most applicable to our case: the same sign di-lepton analysis
\cite{Chatrchyan:2012paa} and two hadronic analyses, namely $\alpha_T$
\cite{Chatrchyan:2013lya} and $\Delta \hat{\phi}$ \cite{Chatrchyan:2013wxa}.
These analyses were performed by CMS based on simplified models with typically only one decay
channel for the gluino, and we re-interpreted their results to set limits on
Yukawa unified SO(10) models. Fortunately, the ATLAS and CMS collaborations provide in many cases the raw data
necessary to reinterpret the searches for new physics in the context of more
elaborate models. E.g.~the model A1 in Ref.~\cite{Chatrchyan:2012paa} assumes
$\mathcal{B}(\widetilde{g}\rightarrow t\bar{t}\widetilde{\chi}^0_1)=100\%$, but
the expected number of SM events in each signal region (cmp.~Tab.~2 in the same
publication) allows us to derive exclusion bounds for our model on the basis of
not having observed a signal. This is in essence what we did in Ref. \cite{Anandakrishnan:2013nca}.
To illustrate our point, in this report, we present a comparison between the bounds obtained from the
$\Delta \hat{\phi}$ analysis for the simplified models T1tttt
and T1bbbb and for the benchmark model YUc. This analysis
was performed by the CMS collaboration with the full dataset (19.4 fb$^{-1}$ at 8 TeV center-of-mass energy)
and rules out gluinos lighter than 1.2 TeV. \\

The strategy of our analysis was the following: we generated 10,000
gluino pair production events with \texttt{PYTHIA}. We switched off all SUSY processes {\em except} $gg
\rightarrow \widetilde{g}\,
\widetilde{g}$ and $q\bar{q}\rightarrow \widetilde{g}\, \widetilde{g}$, since the only
light particles in the spectrum are the gauginos, and the neutralinos and
charginos do not contribute to the event signatures that we will later
consider for the analyses. Moreover, the electroweak processes have much smaller
cross sections and can also be neglected. Following that, the detector simulation was performed using \texttt{Delphes-3.0.9},
with input on the detector type, and efficiency information provided by the collaboration in their analysis. \\

\begin{figure}[h!]
\centering
\subfigure{
\includegraphics[width=0.45\textwidth]{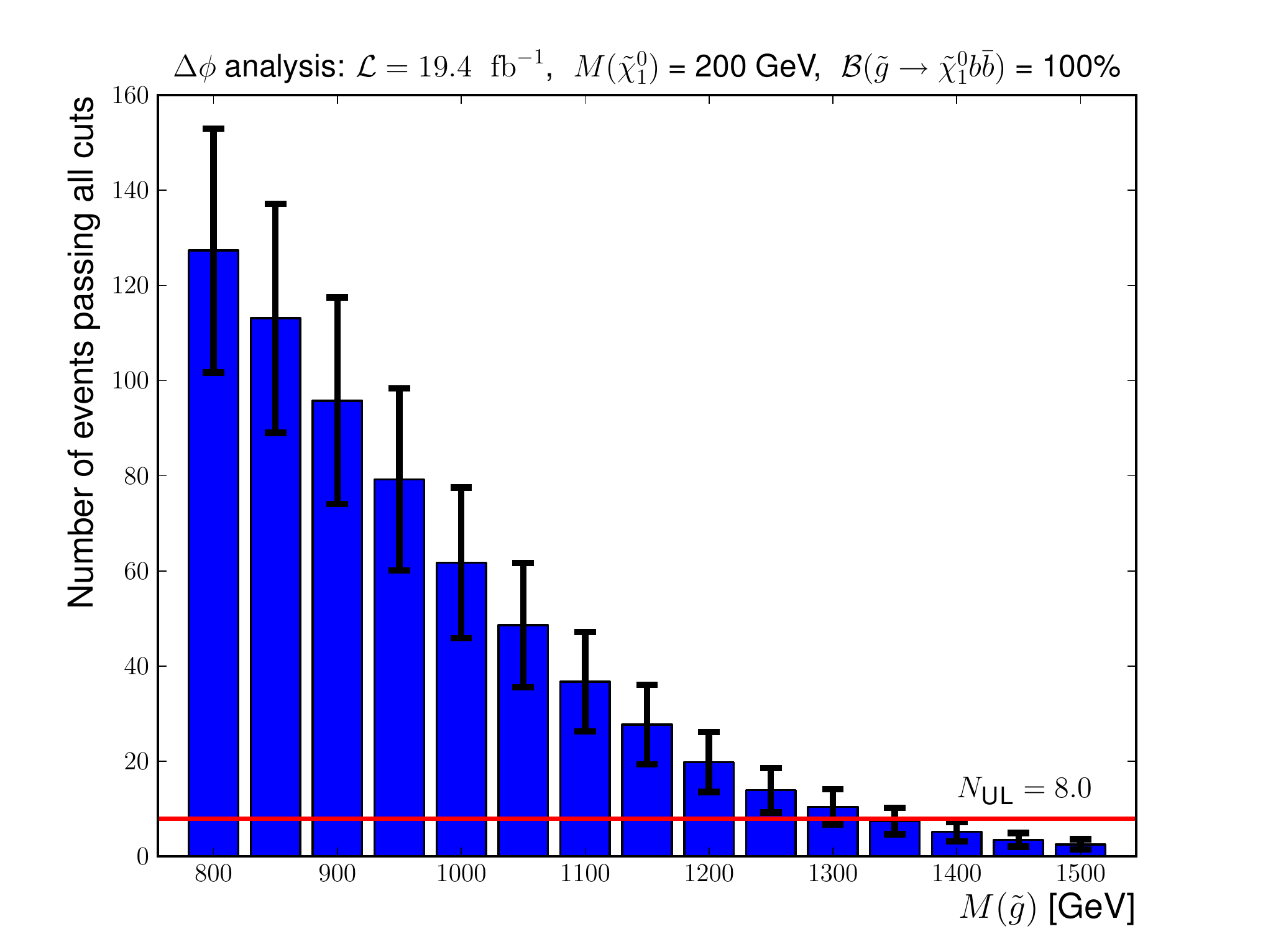}
}
\subfigure{
\includegraphics[width=0.45\textwidth]{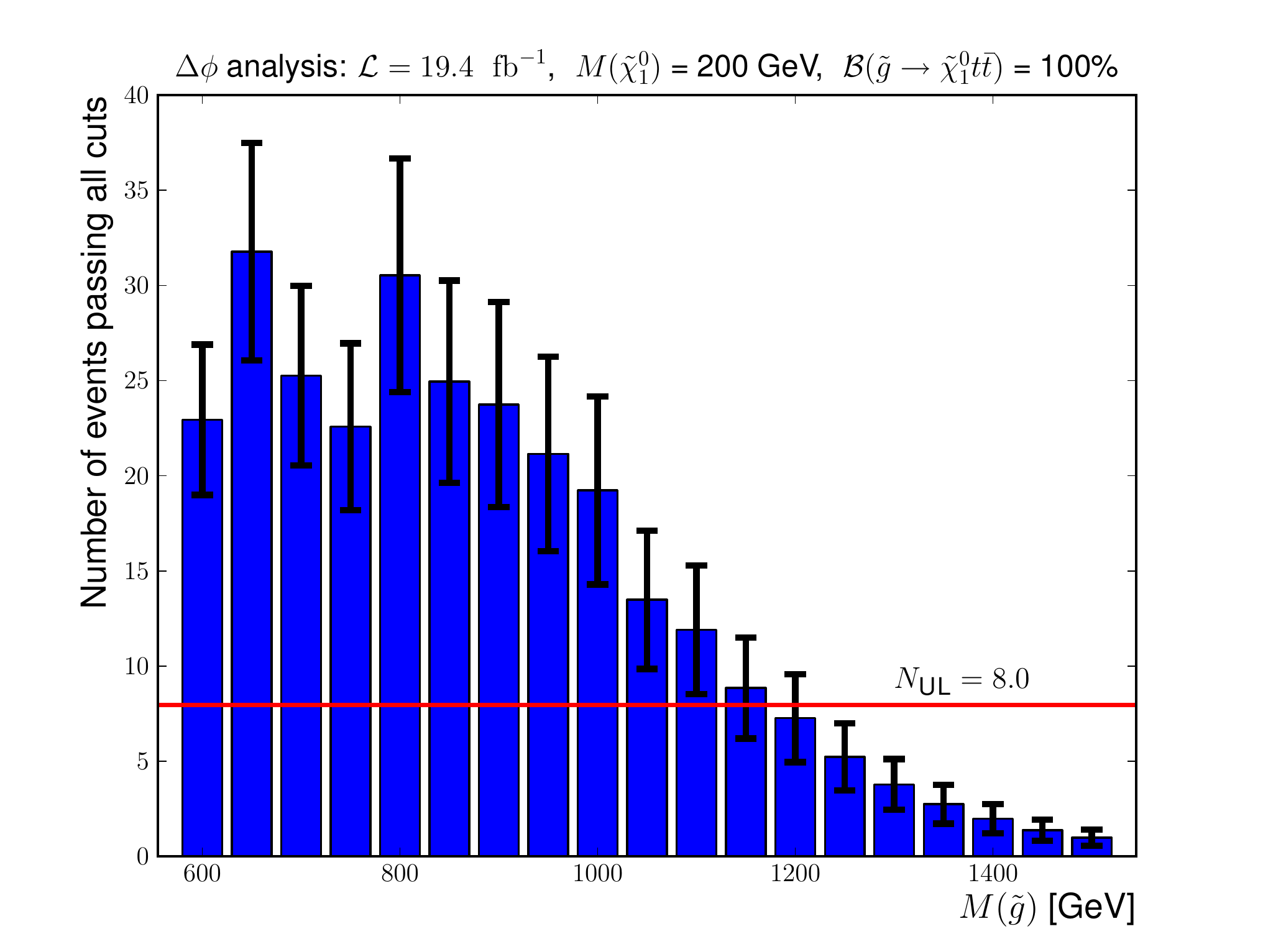}
}
\caption{\footnotesize Validation of CMS analysis \cite{Chatrchyan:2013wxa} for
the simplified SUSY scenarios, $\tilde{g} \rightarrow \tilde{\chi}^0_1 b \bar{b}
$ (left) and
$\tilde{g} \rightarrow \tilde{\chi}^0_1 t \bar{t} $ (right). The CMS analysis
rules out gluinos lighter than 1170 GeV in the T1bbbb model and
1020 GeV in the T1tttt simplified models. Our results are in agreement with the
CMS bounds within the expectations of our numerical tools.}
\label{delphivalidation}
\end{figure}

The $\Delta \hat{\phi}$ analysis \cite{Chatrchyan:2013wxa} looked for events with large transverse missing energy,
jets and b-jets, and with no isolated electrons or muons. This is a typical set of final states
coming from the simplified scenarios T1bbbb and T1tttt. The analysis required at least 3 jets (and at least 1
b-tagged jet)
with $p_T > 50$ GeV and $|\eta| < 2.4$, and binned the events into 14 signal
regions with different ranges of $E_T^{miss}$, $H_T$ and
$N_{\text{b-jet}}$.
The details of the binning can be found in Tab. 2 of Ref.
\cite{Chatrchyan:2013wxa}. For neutralino masses between 0 and 400 GeV, we find the
signal region with $E_T^{miss} > 350$ GeV and $H_T > 1000$ GeV to be the
most constraining one. In addition, events are required to have
$\Delta \hat{\phi}_{\text{min}} > 4.0$, where $ \Delta \hat{\phi}_{\text{min}} =
\text{min} \left( \Delta \phi_i/\sigma_{\Delta \phi_i} \right) $
and $\Delta \phi$ is the angle between a jet and the negative of the
$E_T^{miss}$ vector, and $\sigma_{\Delta \phi_i}$ is the estimated
resolution of $\Delta \phi$. More on this observable can be found in Ref.
\cite{Chatrchyan:2012rg}. QCD background events
are characterized by low $\Delta \phi$, since the $p_T$ mis-measurement gives
rise to most of the missing energy in a QCD event.  By requiring
$\Delta \hat{\phi}_{\text{min}} > 4.0$, most of the QCD backgrounds are
eliminated.\\

The first step was to validate the CMS analysis. CMS quotes that the nominal
b-tagging efficiency is 75\% for jets with a $p_T$ value of 80 GeV. We adapt the Delphes CMS card using the
following information from the CMS analysis: b-tagging efficiency, electron,
and muon isolation criteria, and charged track definition and isolation
criteria. Using the information on the number of observed events and the
expected Standard Model
background from Ref. \cite{Chatrchyan:2013wxa} we determine bounds for the
simplified models T1tttt and T1bbbb. The result is shown in FIG \ref{delphivalidation}. The bounds on the gluino mass obtained from our
validation analysis (CMS analysis) are
1250 (1170) GeV in the T1bbbb model and 1100 (1020) GeV in the T1tttt simplified
model. The agreement between the two quoted numbers is within
the expectations ($\sim$ 20\%) from the tools that we have used.\\

\begin{figure}[h!]
\centering
\includegraphics[width=0.6\textwidth]{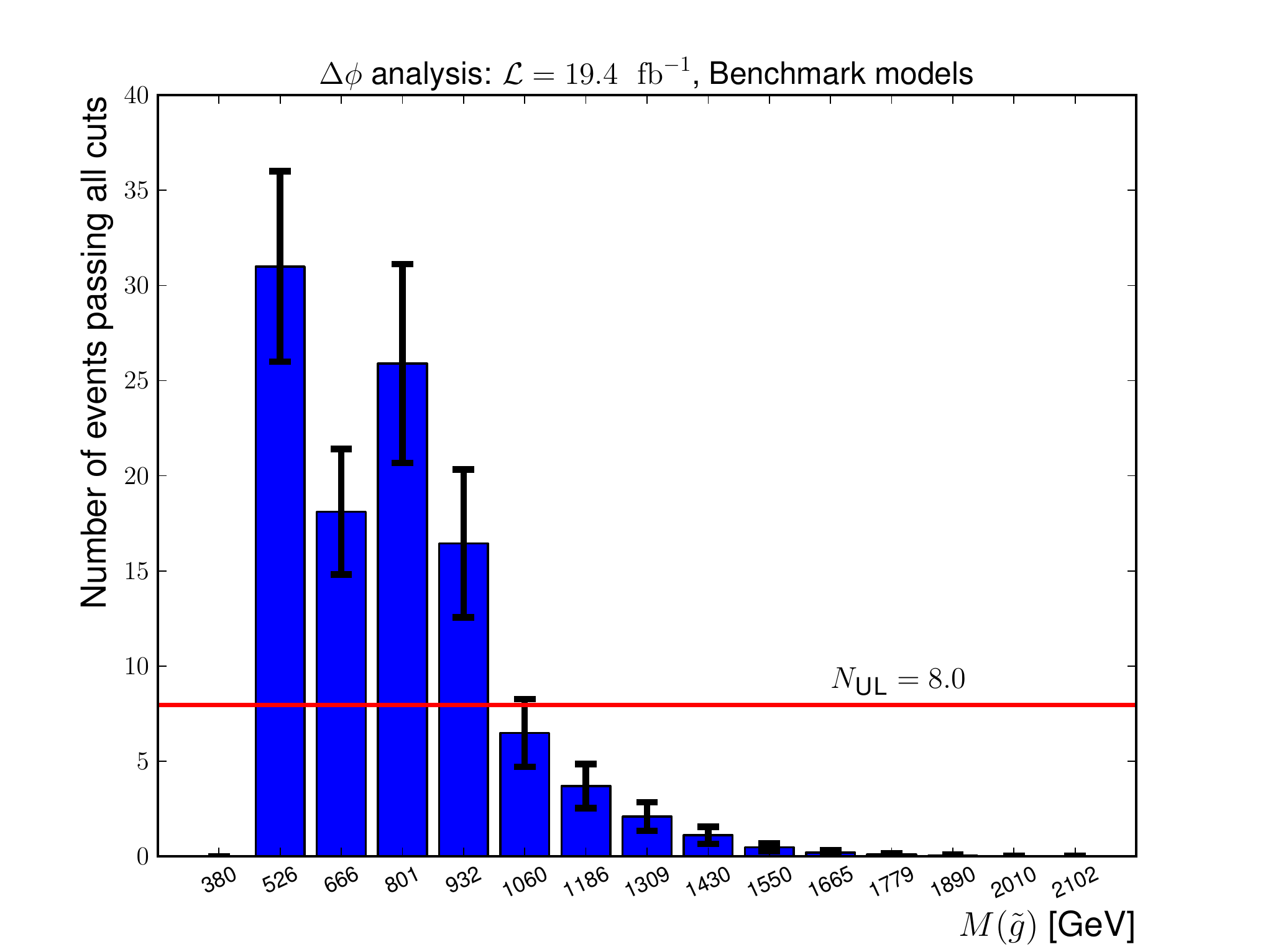}
\caption{\footnotesize The number of events passing all cuts described in
the text for the benchmark models.
The labels on the $x$-axis denote the gluino mass in each benchmark model. Gluino mass of 1060 GeV
represents the benchmark model YUc shown in TABLE \ref{benchmark}. The
red horizontal line shows the  95\% upper limit on the allowed events
from new physics. The error bars are derived from the uncertainty in estimating
the NLO gluino production cross-section.}
\label{fig:CMS-delphi}
\end{figure}

We now proceed to show the results of this analysis, interpreted for the
benchmark models discussed in \cite{Anandakrishnan:2013nca}. We generate 10,000 events for
each of the
benchmark models and apply the same cuts, and since these benchmark models do
not fall into either of the simplified scenarios, we expect the bounds
on the benchmark models to be weaker. In many cases, especially when the mass of
the gluino is greater than 1 TeV (also for the benchmark YUc), loop decays to a gluon and a neutralino are
enhanced and the final states do not have b-tagged jets. The number of events that pass the cuts for
each benchmark model is shown in FIG \ref{fig:CMS-delphi}. It is clear from the
figure that fewer events pass the cuts, since the final states from the gluino decays
do not come from a single decay topology and there are, on average, fewer b-jets.
Thus, in comparison with the simplified models, where we were able to rule out gluinos of at least 1100 GeV (in the T1tttt scenario,
the bound obtained in the T1bbbb scenario was 1250 GeV), we only rule out gluinos
lighter than 1000 GeV with this analysis. The benchmark point YUc is barely ruled out. \\

Our analysis suggests the following: Simplified models are powerful tools for the discovery of new physics at the LHC. However, if and when
the LHC sees a signal, it is unlikely that a simplified model would completely describe it. One
would then have to go beyond the simplified model scenarios. Similarly, the bounds from simplified models may not always provide
an accurate test of more realistic decay scenarios.
We should point out that the best bound we obtained for our benchmark scenarios was from the
same sign di-lepton analysis \cite{Chatrchyan:2012paa} (on which we have not elaborated in this report, for more details see
\cite{Anandakrishnan:2013nca}), where the bounds were as strong as the bounds on the simplified model T1tttt.
Therefore, in principle, existing LHC searches could give stringent bounds on many benchmark scenarios with different decay topologies, but one has to carefully estimate which
is the best search strategy for a given model. \\

This project was partially funded by the US Department of Energy grant DOE/ER/01545-899 and LabEx ENIGMASS. We thank the \emph{Ohio Supercomputer Center} and the \emph{Centre de Calcul de l'Institut National de Physique Nucl\'{e}aire
et Physique des Particules} in Lyon for their computing resources.

\bibliography{snowmass}

\bibliographystyle{utphys}

\end{document}